\documentclass[trackchanges,twocolumn]{aastex701}

\usepackage{graphicx}	
\usepackage{amsmath}	

\begin{document}

\title{The Metallicity Gradient of Sagittarius Dwarf Spheroidal Galaxy Prior to Infall Constrained by S-PLUS  Observations of its Tidal Stream}

\author[0000-0002-8048-8717]{Almeida-Fernandes, F.}
\affiliation{Instituto Nacional de Pesquisas Espaciais, Av. dos Astronautas 1758, Jardim da Granja,12227-010 S\~ao Jos\'e dos Campos, SP, Brazil\\}
\affiliation{Observatorio do Valongo, Universidade Federal do Rio de Janeiro,
Ladeira Pedro Antonio 43, Rio de Janeiro, RJ 20080-090, Brazil}
\email[show]{felipe.fernandes@inpe.br}  

\author[0000-0002-9269-8287]{Limberg, ~G.}
\affiliation{Kavli Institute for Cosmological Physics, University of Chicago, 5640 S Ellis Avenue, Chicago, IL 60637, USA}
\affiliation{Department of Astronomy \& Astrophysics, University of Chicago, 5640 S. Ellis Avenue, Chicago, IL 60637, USA}
\email{}  

\author[0000-0002-0537-4146]{Perottoni,~H.~D.}
\affiliation{Observatório Nacional, Rua General José Cristino, 77, Bairro São Cristóvão, Rio de Janeiro, 20921-400, Brazil \\}
\email{}  

\author[0000-0002-7662-5475]{Amarante,~J.~A.~S.}
\affiliation{Department of Astronomy, School of Physics and Astronomy, \\ Shanghai Jiao Tong University, 800 Dongchuan Road, Shanghai 200240, China}
\affiliation{State Key Laboratory of Dark Matter Physics, School of Physics and Astronomy, \\ Shanghai Jiao Tong University, Shanghai 200240, China }
\email[show]{joaoant@gmail.com}

\author[0000-0001-7057-9685]{Bolutavicius,~G.~F.}
\affil{Universidade de S\~ao Paulo, Instituto de Astronomia, Geof\'isica e Ci\^encias Atmosf\'ericas, \\ Departamento de Astronomia, SP 05508-090, S\~ao Paulo, Brazil}
\email{}  

\author[0000-0001-9079-9511]{Cordeiro,~V.}
\affiliation{Observatório Nacional, Rua General José Cristino, 77, Bairro São Cristóvão, Rio de Janeiro, 20921-400, Brazil \\}
\email{}  

\author[0000-0001-5740-2914]{Borges Fernandes,~M.}
\affiliation{Observatório Nacional, Rua General José Cristino, 77, Bairro São Cristóvão, Rio de Janeiro, 20921-400, Brazil \\}
\email{}  

\author[0000-0003-4479-1265]{Placco,~V.~M.}
\affiliation{NSF NOIRLab, Tucson, AZ 85719, USA}
\email{}  

\author[0000-0002-7758-656X]{da Silva,~A.~R.}
\affil{Universidade de S\~ao Paulo, Instituto de Astronomia, Geof\'isica e Ci\^encias Atmosf\'ericas, \\ Departamento de Astronomia, SP 05508-090, S\~ao Paulo, Brazil}
\email{}

\author[0000-0001-8789-6230]{Santos-Silva,~T.}
\affiliation{Universidade Estadual de Feira de Santana, Departamento de F\'isica, Feira de Santana, BA 44.036-900, Brasil}
\affil{Universidade de S\~ao Paulo, Instituto de Astronomia, Geof\'isica e Ci\^encias Atmosf\'ericas, \\ Departamento de Astronomia, SP 05508-090, S\~ao Paulo, Brazil}
\email{}  

\author[0000-0001-6987-1531]{Machado-Pereira,~E.}
\affiliation{Observatório Nacional, Rua General José Cristino, 77, Bairro São Cristóvão, Rio de Janeiro, 20921-400, Brazil \\}
\email{}  

\author[0000-0002-7529-1442]{Santucci,~R.~M.}
\affiliation{Universidade Federal de Goiás, Instituto de Estudos Socioambientais, Planetário, Goiânia, GO 74055-140, Brazil}
\affiliation{Universidade Federal de Goiás, Campus Samambaia, Instituto de Física, Goiânia, GO 74001-970, Brazil}
\email{}  

\author[0000-0003-3153-5123]{Menéndez-Delmestre,~K.}
\affiliation{Observatorio do Valongo, Universidade Federal do Rio de Janeiro,
Ladeira Pedro Antonio 43, Rio de Janeiro, RJ 20080-090, Brazil}
\email{}  

\author[0000-0003-2374-366X]{Gonçalves,~T.~S.}
\affiliation{Observatorio do Valongo, Universidade Federal do Rio de Janeiro,
Ladeira Pedro Antonio 43, Rio de Janeiro, RJ 20080-090, Brazil}
\email{}  

\author[0000-0001-7479-5756]{Rossi,~S.}
\affil{Universidade de S\~ao Paulo, Instituto de Astronomia, Geof\'isica e Ci\^encias Atmosf\'ericas, \\ Departamento de Astronomia, SP 05508-090, S\~ao Paulo, Brazil}
\email{}  

\author[0000-0002-2484-7551]{Kanaan,~A.}
\affiliation{Universidade Federal de Santa Catarina, Campus Universitário Reitor João David Ferreira Lima, Florianópolis, 88040-900, Brazil \\}
\email{}  

\author[0000-0002-4064-7234]{Schoenell,~W.}
\affiliation{The Observatories of the Carnegie Institution for Science, 813 Santa Barbara St., Pasadena, CA91101, USA \\}
\email{}  

\author[0000-0002-0138-1365]{Ribeiro,~T.}
\affiliation{Rubin Observatory Project Office, 950 N. Cherry Ave, Tucson, 85719, US\\}
\email{}  

\author[0000-0002-5267-9065]{Mendes de Oliveira,~C.}
\affil{Universidade de S\~ao Paulo, Instituto de Astronomia, Geof\'isica e Ci\^encias Atmosf\'ericas, \\ Departamento de Astronomia, SP 05508-090, S\~ao Paulo, Brazil}
\email{}  

\begin{abstract}

We study the metallicity distribution along the Sagittarius (Sgr) stream using photometric metallicities from S-PLUS DR4, combined with Gaia DR3 kinematics and APOGEE DR17 spectroscopy. Our analysis confirms that the leading arm (Galactic latitude $b > 0$) is systematically more metal-poor than the trailing arm ($b < 0$)  by 0.15–-0.20 dex, and reveals a clear negative metallicity gradient along the leading arm. The trailing arm shows no significant overall gradient but displays distinct inner (negative) and outer (positive) trends. These features are consistently recovered across different photometric estimators and agree with spectroscopic data. We compare these results with predictions from an $N$-body simulation, in which metallicities were assigned according to a set of imprinted radial gradients in the progenitor. We were able to constrain the original metallicity gradient of the Sgr progenitor to be between $-0.38$ and $-0.24$ dex kpc$^{-1}$ based on photometric data, and $-0.42$ to $-0.10$ dex kpc$^{-1}$ from APOGEE. These values are consistent with gradients observed in other Local Group dwarf galaxies. Our findings demonstrate that metallicity-sensitive photometric surveys such as S-PLUS are powerful tools for reconstructing the chemodynamical evolution of disrupted satellites.

\end{abstract}



\section{Introduction} \label{sec:intro}

The Sagittarius (Sgr) dwarf spheroidal (dSph) galaxy is one of the most massive satellites of the Milky Way (stellar mass $M_\star \approx 3\times 10^7 M_\odot$\footnote{Assuming this galaxy's $V$-band absolute magnitude $M_V = -13.27$ \citep[][]{Majewski+2003}, $M_{V,\odot} = +4.81$ for the Sun \citep[][]{Willmer2018}, and a mass-to-light ratio of 2, which is appropriate for non-star-forming systems \citep[e.g.,][]{Simon2019}.}) as well as one of the closest to the Sun \citep[e.g.,][$\sim$26.5\,kpc]{McConnachie2012, Pace2024}. In the sky, Sgr dSph is located in the direction of the Galactic center, behind the Galactic bulge. The substantial amount of foreground contamination, as well as high extinction, prevented earlier discovery of this satellite despite its mass and proximity. The Sagittarius dwarf galaxy was eventually uncovered by \citet{Ibata+1994}, who identified a group of comoving stars in this region. The most intriguing aspect of Sgr dSph is, nevertheless, the existence of an extensive stellar stream associated with it, the so-called ``Sgr stream'' \citep{Mateo1998, Ibata2001sgrStream, Majewski+2003, Yanny2009sgr}, which is produced by severe tidal stripping experienced by the system beginning around 6 billion years ago \citep[see the simulation work by, e.g.,][]{Helmi2001sgr, Law2005, Law+Majewski2010, VeraCiro+Helmi2013, Gomez+2015, Dierickx2017sgr, Fardal2019sgr, Vasiliev+2021a, Wang_HF+2022}.

As galaxy-galaxy interactions are ubiquitous phenomena within concordance cosmology \citep{FaberGallagher1979, Kauffmann+1993, Springel2006}, the Sgr system \citep[stream$+$core;][]{Vasiliev+2021a} provides the best available environment for understanding the influence of a massive host (i.e., the Milky Way) on a satellite. In this context, stellar population variations have been identified between the stream (characterized by lower metallicity) and the main body (characterized by higher metallicity) of Sgr dSph \citep[][]{Chou+2007, Monaco+2007}. More surprisingly, there is an [Fe/H]\footnote{${\rm[X/Y]} = \log{(N_{\rm X}/N_{\rm Y})_\star} - \log{(N_{\rm X}/N_{\rm Y})_\odot}$, where $N_{\rm X}$ ($N_{\rm Y}$) is the number density of atoms of element X (Y) for a given star ($\star$) relative to the Sun ($\odot$).} difference of ${\sim}0.3$\,dex between the \textit{leading} (Galactic latitude $b>0^{\circ}$) and \textit{trailing} ($b<0^{\circ}$) arms of the stream \citep{Monaco+2007, JingLi2016sgr, JingLi2019sgr, Yang2019sgrLAMOST, Ramos+2022}. These observations are often attributed to star-formation bursts induced during pericentric passages of Sgr dSph around the Milky Way \citep[][]{Siegel2007SFHsgr, Hasselquist2021dwarf_gals}. However, a self-consistent chemodynamical model of this interplay still does not exist.

Apart from these stellar population differences between the remaining core of Sgr and its leading and trailing arms, many authors have reported chemical variations within the stream itself \citep{Bellazzini2006sgr, Chou2010sgr, Keller+2010, Hyde+2015}. \citet{Hayes+2020} presented a recent measurement of metallicity gradients along both arms of the stream by employing high-resolution ($\mathcal{R} > 20{,}000$) spectroscopic data from the Apache Point Observatory Galactic Evolution Experiment \citep[APOGEE;][]{Majewski+2017, Ahumada+2020}, including 166/710 stars from the stream/core. They found mild internal metallicity gradients along each arm\footnote{This quantity is computed in a coordinate system based on Sgr's orbital plane \citep[][see Section \ref{sec:data}]{Majewski+2003}.}, which become clearer when anchoring the gradient to the metallicity of the main body. While the difference between the core and the tidal arms can be explained by bursts of star formation in the core induced by pericentric passages, we expect no such bursts to occur within the arms themselves. Therefore, the metallicity variations observed along the arms should reflect the progenitor's pre-existing radial gradient.

More recently, thanks to the availability of increasingly larger samples of reliable Sgr members, in particular in the very metal-poor regime ($\rm[Fe/H] \lesssim -2$), it has been demonstrated that the stream contains both a metal-rich, dynamically colder (i.e. lower velocity dispersion) population as well as a metal-poor, dynamically hotter component \citep[][see also \citealt{Gibbons2017sgr}]{Johnson2020sgr}. Most crucially, \citet[][]{Limberg+2023} has shown, by comparison with a state-of-the-art $N$-body model of the Sgr system \citep{Vasiliev+2021a}, that the existence of these two components is naturally explained if a metallicity gradient was already in place in Sgr \textit{prior} to its infall. If confirmed, this would carry important implications for our galaxy formation understanding, since it would show that classical dwarf galaxies can develop metallicity gradients on relatively short timescales, already by redshift $\sim$0.5.

This work builds upon these recent discoveries regarding the Sgr stream. In particular, our goal is to falsify (or confirm) the connection between the present-day metallicity variations on the stream and the hypothesized metallicity gradient in Sgr dSph prior to infall. For this task, we consolidated a large sample of ($\gtrsim$ 5000) stream members with available photometric metallicities derived from narrow-band imaging from the Southern Photometric Local Universe Survey \citep[S-PLUS][]{MendesdeOliveira+2019}. Well-calibrated photometric metallicities have the advantage of being spatially unbiased, providing realistic information on stellar-population trends within dwarf galaxies \citep[e.g.,][]{Barbosa+2025}. Finally, we aim to reconstruct the original metallicity gradient of Sgr from our observations of the stream as conjectured to be possible by \citet[][]{Limberg+2023}. This is achieved by painting metallicity gradients onto an $N$-body simulation, as previously explored by \citet{Cunningham+2024}, who used metallicities inferred from Gaia DR3 \citep[][]{GaiaCollaboration+2023} Blue/Red Photometer spectra \citep{Andrae+2023}. Their analysis allowed the authors to constrain the original radial metallicity gradient of the progenitor galaxy. While their study concentrated on broad trends, here we provide a more quantitative reconstruction of the progenitor's metallicity gradient based on the S-PLUS photometric metallicities.

The paper is organized as follows. In Section \ref{sec:data} we present the data that were used for our analyses. Our results for the metallicity distribution and metallicity gradients are presented in Section \ref{sec:gradient}. In Section \ref{sec:nbody} we compare our previous results to the predictions of N-body simulations in order to infer the original properties of the Sgr progenitor galaxy. Finally, in Section \ref{sec:conclusions} we present our conclusions.
\section{Data} \label{sec:data}

This section describes the data used in our analyses. We first present the sample of Sgr stream stars identified by \citet{Ramos+2022} from Gaia data, followed by the S-PLUS DR4 \citep{Herpich+2024} catalog and the description of the machine learning models used to estimate stellar metallicities from photometry.

\subsection{Members of Sgr stream in the Gaia era}

The Gaia survey \citep{GaiaCollaboration+2016} has played a crucial role in the all-sky mapping of the Sgr stream, thanks to its precise and extensive astrometric data. The Sgr stream presents unique challenges as it is expected to have undergone multiple pericenter passages in its orbit trajectory \citep[e.g.][]{Purcell+2011, Laporte+2018} and its stars are distributed over distances of 20 to 100 kpc \citep[e.g.][]{hernitschek+2017}. 

\begin{figure*}[!ht]
    \begin{center}
        \includegraphics[width=\linewidth, angle=0]{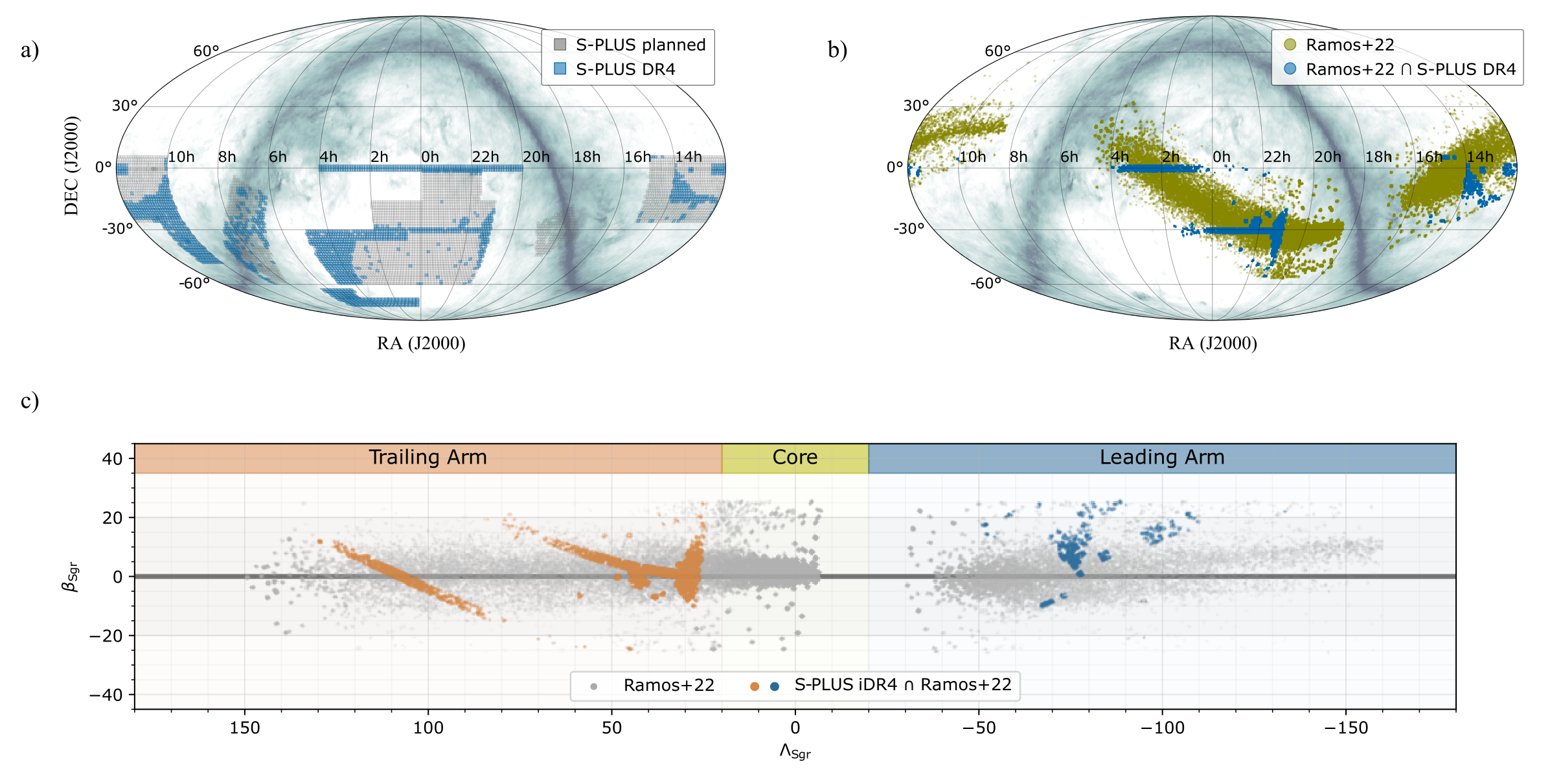}
        \caption{a) Aitoff projection in Equatorial coordinates showing the coverage of the S-PLUS survey (DR4) indicated by the blue region. b) spatial distribution of the R+22 sample (yellow) and the crossmatch between the S-PLUS DR4 data and the R+22 sample (blue). c) Distribution of stars in the Sgr stream coordinate system, where we define the trailing arm as the region $20^{\circ} < \Lambda_{\mathrm{Sgr}} \leq 180^{\circ}$ (orange), the leading arm as the region $-180^{\circ} < \Lambda_{\mathrm{Sgr}} < -20^{\circ}$ (blue), and the core as the region $|\Lambda_{\mathrm{Sgr}}| < 20^{\circ}$ (yellow).}\label{fig:footprint}
    \end{center}
\end{figure*}

\citet[][hereafter, R+22]{Ramos+2022} used Gaia DR3 \citep{GaiaCollaboration+2021} and significantly increased the sample of Sgr stream candidates in comparison to previous studies \citep{Antoja+2020, Ramos+2020, Ibata2020}. Using their final sample, containing $700\,000$ candidate stars, R+22 were able to map the Sgr stream with unprecedented detail and detect bifurcations of the stream in both hemispheres. We adopt the sample of \citet{Ramos+2022} as the reference for the stellar data belonging to the Sgr stream. The sky distribution of the Sgr candidate stars is shown in yellow in the right panel of Figure \ref{fig:footprint}.

\subsection{S-PLUS DR4}

The photometric sample used in this work was obtained from the S-PLUS data release 4 \citep{Herpich+2024}, which covers about 3000 square degrees of the southern sky in 12 filters, including 7 narrow bands \citep{MendesdeOliveira+2019, Cenarro+2019} and 5 broad-bands (the Javalambre $u$-band and SDSS-like $g$, $r$, $i$ and $z$ filters), reaching photometric depths between 19.7 and 21.5 mag\footnote{S-PLUS adopts the AB magnitude system \citep{Almeida-Fernandes+2022}.} for a minimum signal-to-noise ratio $S/N = 3$. The seven narrow-band filters of the system are strategically placed to trace prominent spectral features. Specifically, they are centered on \ion{O}{2} (J0378)\footnote{The Javalambre narrow-band filters used in S-PLUS are named according to the central wavelength of the filter in units of nm.}, the \ion{Ca}{2} H and K doublet (J0395), the H$\delta$ Balmer line (J0410), the CH G-band (J0430), the Mg-b triplet (J0515), the H-alpha (J0660), and the \ion{Ca}{2} infrared triplet (J0861). This configuration ensures the sensitivity of S-PLUS to stellar atmospheric parameters and metallicity beyond what is possible with broad-band photometry alone. The footprint of S-PLUS DR4 is shown, in blue, in the left panel of Figure \ref{fig:footprint}. 

S-PLUS includes observations in both Galactic hemispheres, covering a significant portion of the trailing arm in the southern Galactic hemisphere, and also a less extended region of the leading arm in the northern Galactic hemisphere. This coverage is shown in the right panel of Figure \ref{fig:footprint}, where the crossmatch between S-PLUS DR4 and the R+22 sample is represented in blue. The Sgr core is not covered by this S-PLUS data release and will be explored in a future study.

In the bottom panel of Figure \ref{fig:footprint}, we show the distribution of the stars in the Sgr stream coordinate system. In this work, we adopt the transformation of \citet{Majewski+2003} and define the trailing arm as the region $20^{\circ} < \Lambda_{\mathrm{Sgr}} \leq 180^{\circ}$, and the leading arm as the region $-180^{\circ} < \Lambda_{\mathrm{Sgr}} < -20^{\circ}$\footnote{We adopt the conventional range to define the leading and trailing arms, even though our observed sample spans only a subset of this. This choice allows us to keep the same definition for both observations and simulations.}.  

\subsection{Photometric metallicities}

The presence of narrow-band filters in S-PLUS, centered on key spectral features, enables its use for estimating stellar metallicities. Filters such as J0395, J0410, J0515, and J0861 are particularly important in constraining [Fe/H] values for FGK-type stars. This capability has already been demonstrated in several S-PLUS studies, including the identification of metal-poor stars \citep{Placco+2022, Almeida-Fernandes+2023, Perottoni+2024}, confirmed with high-resolution spectroscopic follow-ups \citep{Placco+2021, Placco+2023}, and the characterization of chemical abundances via machine learning techniques \citep{Whitten+2021, Molina-Jorquera+2024,  Quispe-Huaynasi+2024, FerreiraLopes+2025}. Similar methodologies have also been applied to the Javalambre Photometric Local Universe Survey (J-PLUS) survey \citep{Cenarro+2019}, which shares the same instrumental setup as S-PLUS, and has likewise yielded successful estimates of stellar metallicities and atmospheric parameters \citep{Whitten+2019, Galarza+2022, Yang+2022, Wang+2022, Quispe-Huaynasi+2023, Huang+2024}.

In this work, we adopt photometric metallicities derived from S-PLUS DR4 \citep{Herpich+2024}
via two machine learning techniques: the Random Forest (RF) method is used as our primary source of [Fe/H] estimates, while the neural network model (NN) provides an independent validation. As demonstrated in Appendix \ref{apend:photmet}, the photometric metallicities are not reliable for cool stars ($g-i > 1.7$). As a result, these stars were removed from our final sample, corresponding to the removal of 92 (1\%) of the stars in the RF sample, and 456 (3\%) for the ANN sample. The methods are described below.

\subsubsection{Random Forest (RF)}

The estimation of $T_{\rm eff}$, $\log{(g)}$, and [Fe/H] with a Random Forest (RF) approach has been described in \citet{Silva2023}. Here we provide a brief summary. The models were developed using Machine Learning algorithms based on Random Forest, trained on a cross-matched sample of S-PLUS DR4 photometry, stellar parameters from the  Large Sky Area Multi-Object Fiber Spectroscopic Telescope \citep[LAMOST,][]{Cui+2012} DR8 catalog, and distances from \citet{Bailer-Jones+2021}. The final training set consisted of 51{,}278 stars, of which 42{,}501 had reliable parallaxes from Gaia (\texttt{parallax\_over\_error} $>= 5$ and \texttt{ruwe} $<= 1.4$). The predictions of $T_{\rm eff}$ and [Fe/H] for each object were made from its apparent magnitudes, and the prediction of $\log{(g)}$ was made from either its apparent magnitudes (for objects without reliable parallaxes) or from its absolute magnitudes (for objects with reliable parallaxes). 

\subsubsection{Artificial Neural Network (ANN)}

Additionally, we also considered the photometric metallicity estimates from an artificial neural network model, provided as a value-added catalog in S-PLUS DR4. The method follows the procedures described in \citet{Whitten+2021} for the S-PLUS DR2, and was retrained for the S-PLUS DR4 dataset.

In Bolutavicius et al. (in prep), we conducted a validation of the photometric metallicities by comparing the estimates to the spectroscopic values from LAMOST, APOGEE, and the Galactic Archaeology with HERMES \citep[GALAH,][]{DeSilva+2015} surveys. After excluding stars cooler than 4250 K, the results indicate a median bias of $-$0.046, $-$0.041, and $-$0.231 dex, and a standard deviation of $-$0.244, $-$0.209, and $-$0.315 dex, respectively, for each survey (for metallicities below $-$0.8~dex). A comparison between our photometric metallicities and those from APOGEE DR17 is also shown in Appendix~\ref{apend:photmet}.

\subsection{APOGEE}

In order to compare the metallicity gradients obtained for the Sgr stream using S-PLUS with those from \citet{Hayes+2020}, we selected stars from APOGEE data release 17 (DR17; \citealt{APOGEEdr17}) spectroscopic catalog. We applied several criteria to filter the APOGEE DR17 sources, including spectroscopic flags ({\tt ASPCAPFLAG == 0} and {\tt STARFLAG == 0}; see \citealt{Jonsson2020}), reliable estimates of [Fe/H]  (i.e., {\tt FE\_H\_FLAG == 0}), and $S/N > 50$ pixel$^{-1}$. Additionally, we implemented cuts to select only giant stars, with effective temperature limits ($3500 < T_{\rm eff}/{\rm K} < 6000$) and $\log{(g)} < 3.5$. We also used \citet{VasilievBaumgardt2021gcs}'s catalog and excluded stars that are likely associated with globular clusters using the same procedure as \citet[][]{Limberg+2022}.

Subsequently, the APOGEE sample was cross-matched with the Gaia DR3 catalog, using a $1.5\arcsec$ search radius, to obtain parallaxes and proper motions. Then, in order to obtain the Sgr stars (stream$+$dSph) from APOGEE, we simply adopt the same members as \citet[][]{Hayes+2020}, which will guarantee the consistency of our analysis with these authors. Therefore, we are left with the same Sgr members as \citet[][]{Hayes+2020}, but with updated metallicities from  APOGEE DR17.

\section{Metallicity of the Sagittarius Stream}
\label{sec:gradient}

In this Section, we analyze the photometric metallicities derived for the stars in the Sgr stream. First, we compare the metallicity distribution of the trailing and leading arms, followed by the analyses of the metallicity gradients measured along both arms and their comparison to the gradient measured from spectroscopic metallicities.

\subsection{Metallicity distribution function}

\begin{figure}
 \includegraphics[width=\columnwidth]{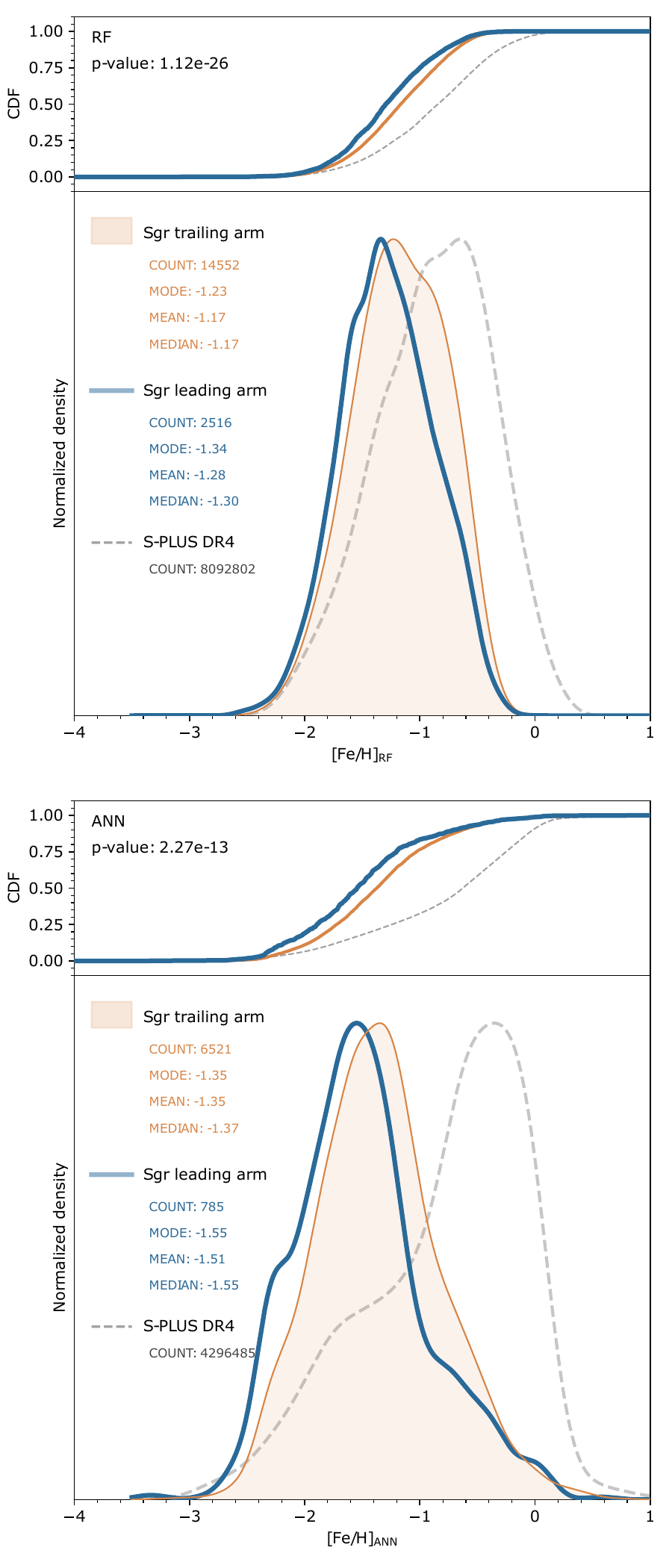}
 \caption{Metallicity distribution functions (MDFs) of photometrically selected Sagittarius stream members, based on the ANN (top) and RF (bottom) metallicity estimates. Each panel includes a cumulative distribution function (CDF, upper subpanel) and a kernel density estimate (KDE, lower subpanel) for stars in the leading (blue) and trailing (orange) arms. The CDF plot also displays the p-value calculated from the KS-test. For reference, the normalized MDF of the full S-PLUS DR4 sample is shown in gray.}
 \label{fig:mdf-ann}
\end{figure}

In Figure \ref{fig:mdf-ann}, we show the metallicity distribution functions (MDFs) of the selected Sgr stream members based on the ANN (top) and RF (bottom) metallicity estimates. In each case, we show the cumulative distribution function (upper subpanel) and a kernel density estimate of the normalized MDF (lower subpanel). For comparison, the overall S-PLUS DR4 MDF is shown in gray. As expected, Sgr stars populate the metal-poor tail of the S-PLUS distribution.

For both metallicity estimates, the trailing arm exhibits a higher fraction of metal-rich stars ([Fe/H]~$\geq~-1$). In contrast, the leading arm shows an enhanced number of very metal-poor stars ([Fe/H]~$\leq~-2$), which is especially evident in the ANN estimates. These trends are consistent with spectroscopic studies that have reported systematically lower metallicities in the leading arm relative to the trailing arm \citep[e.g.][]{Carlin2018sgr, Hayes+2020, Limberg+2023}.

Quantitatively, we find that the leading arm is 0.10–0.20 dex more metal-poor than the trailing arm. This difference agrees with the $\sim$0.2 dex offset reported by \citet{Limberg+2023}. Notably, the median metallicities derived from ANN estimates are slightly more metal-poor than those reported in \citeauthor{Limberg+2023} by 0.08 dex (leading) and 0.09 dex (trailing), while the RF values are  more metal-rich by 0.17 dex and 0.12 dex, respectively. 

Additionally, we performed a Kolmogorov-Smirnov test to evaluate the hypothesis that the two distributions originate from the same parent population. The resulting p-values of $2.27\times10^{-13}$ (ANN) and $1.12\times10^{-26}$ (RF) are both far below any conventional significance threshold, which allows us to reject the null hypothesis. This strongly supports the conclusion that the metallicity distributions of the two arms are intrinsically different. This result is consistent for both sets of photometric metallicities.

The recovery of this established characteristic of the Sgr stream, using just photometric metallicities, demonstrates its efficacy for studying stellar populations. It is important to note that our sample comprises approximately 17068 (RF) and 7327 (ANN) stars (about an order of magnitude larger than most previous spectroscopic analyses), a fact which allows for more robust statistical comparisons between the two arms. The differences in sample sizes between both methods are due to different requirements of data quality to produce a reliable output, with the ANN method being restricted to brighter sources (the median $g$ magnitude for the RF sample is 18.9, in contrast to a median of 17.9 for the ANN sample). Overall, our findings reinforce the established picture that the Sgr leading arm is systematically more metal-poor than the trailing arm, which could be explained by the scenario in which the stars forming the leading arm were stripped earlier and corresponds to the more metal-poor outer regions of the progenitor.


\subsection{Observed metallicity gradient}

\begin{figure*}
 \includegraphics[width=\linewidth]{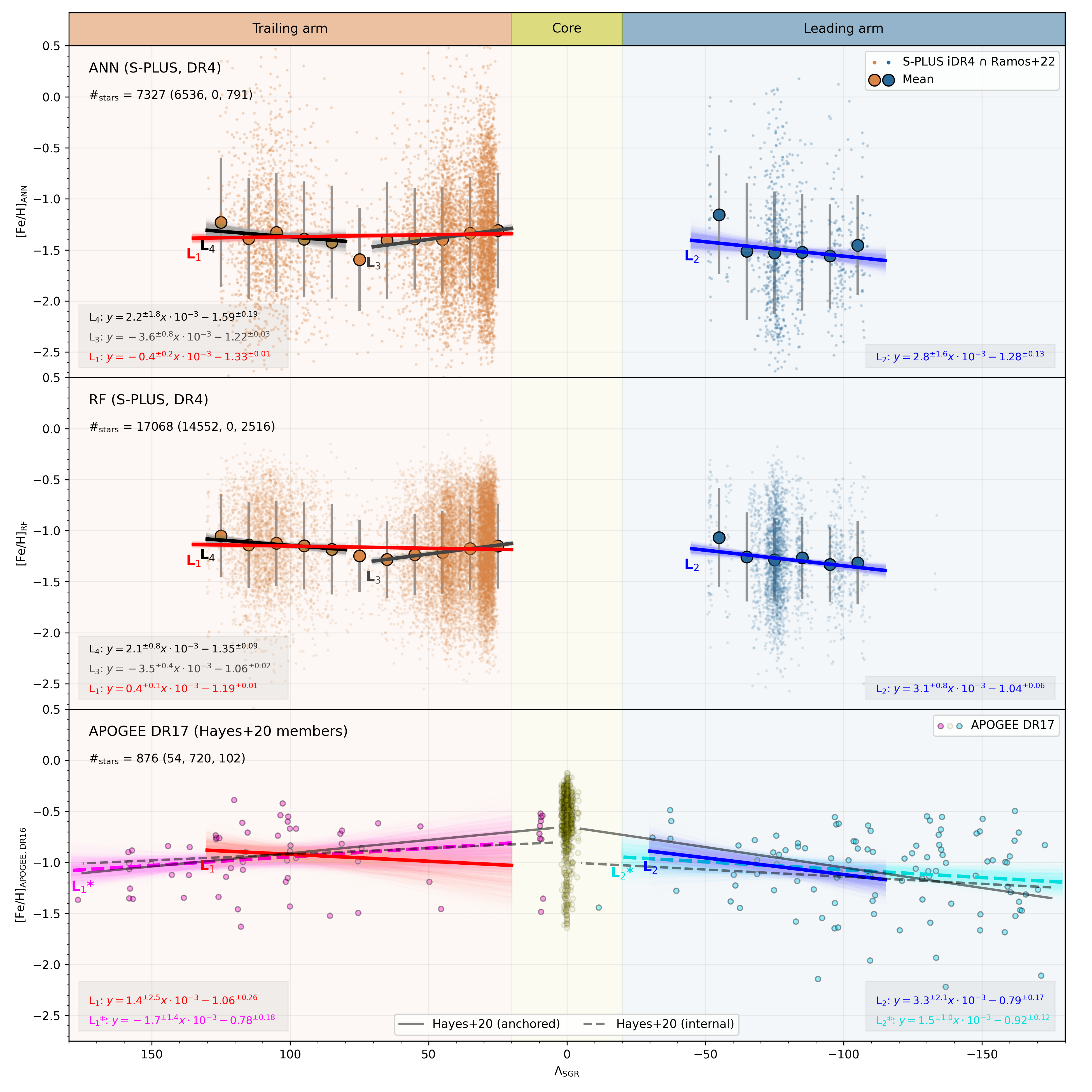}
 \caption{Metallicity distribution as a function of the coordinate $\Lambda_{\mathrm{SGR}}$ for stars in the Sagittarius stream, considering photometric metallicities derived via ANN (top panel) and RF (middle panel) methods from S-PLUS DR4, and spectroscopic metallicities from APOGEE DR16 (bottom panel) for the \citet{Hayes+2020} Sgr members. The trailing arm ($\Lambda_{\mathrm{SGR}} > 20^\circ$) is shown in red, the leading arm ($\Lambda_{\mathrm{SGR}} < -20^\circ$) in blue, and the Sagittarius core ($-20^\circ \leq \Lambda_{\mathrm{SGR}} \leq 20^\circ$) in yellow (APOGEE only). Large circles represent mean metallicities computed in 10$^\circ$ bins, with vertical lines indicating metallicity dispersion. Linear fits to the metallicity gradients are indicated with labeled lines, with shaded regions depicting bootstrap uncertainties. For comparison, internal (gray dashed) and anchored (solid black) metallicity gradients from \citet{Hayes+2020} are shown in the bottom panel.}
 \label{fig:gradient_splus}
\end{figure*}

The median metallicities derived for both the leading and trailing arms are significantly lower than those reported for the Sgr core from spectroscopic studies, which typically find mean metallicities in the range $-0.6$~to~$-0.4$~dex \citep[e.g.,][]{Monaco2005sgr, Bellazzini+2008, Carretta+2010}. This finding supports the widely accepted scenario in which stars in the outskirts of the progenitor system were the first to be stripped in the earlier pericentric passages, as these stars have lower bound energies and are expected to be more metal-poor if we assume there is a radial metallicity gradient in the satellite before the merger. Consequently, a natural outcome of this process is the emergence of a metallicity gradient along the stream, with stars located further from the core expected to be more metal-poor on average.

Observational evidence for such metallicity gradients has been reported in multiple spectroscopic studies. Early works using M giants \citep{Chou+2007} and red giant branch stars (RGB) \citep{Monaco+2007} observed a clear decline in [Fe/H] along the leading tail, which was extensively confirmed in subsequent studies \citep[e.g.][]{Keller+2010, Carlin+2012, Shi+2012, Hyde+2015, Hayes+2020}. Recently, \citet{Cunningham+2024} derived metallicities for 34{,}240 RGB stars using Gaia BP/RP spectra, identifying metallicity gradients both along and across the stream. Additionally, \citet{Muraveva+2025} analyzed RR Lyrae stars from Gaia DR3, confirming a significant gradient in the leading arm but a negligible one in the trailing arm.

\begin{table}[]
\centering
\caption{Summary of the estimated gradients considering different methods and regions for the trailing and leading arms of the Sgr stream. \citep{Hayes+2020} results are shown for comparison. For both arms, negative values mean that metallicity decreases when the distance from the core increases.}
\label{tab:gradient_summary}
\begin{tabular}{lcc}
\hline \hline
\multicolumn{3}{c}{Trailing Arm}                                            \\ \hline
                    & $\Lambda_\mathrm{Sgr}$ & Slope                     \\
Method              & $\in$                  & $\times 10^3$ dex deg$^{-1}$ \\ \hline
S-PLUS DR4 ANN      & {[}20, 130{]}          & $-0.4 \pm 0.2$               \\
S-PLUS DR4 RF       & {[}20, 130{]}          & $0.4 \pm 0.1$                \\
APOGEE DR17         & {[}20, 130{]}          & $+1.4 \pm 2.5$               \\
APOGEE DR17         & {[}20, 180{]}          & $-1.7 \pm 1.4$               \\
Hayes+20 (anchored) &           & $-2.6 \pm 0.4$               \\
Hayes+20 (internal) &           & $-1.2 \pm 0.9$               \\ \hline \hline
\multicolumn{3}{c}{Leading Arm}                                             \\ \hline 
                    & $\Lambda_\mathrm{Sgr}$ & Slope                     \\
Method              & $\in$                  & $\times 10^3$ dex deg$^{-1}$ \\ \hline
S-PLUS DR4 ANN      & ${[}-115, -20{]}$          & $-2.8 \pm 1.6$               \\
S-PLUS DR4 RF       & ${[}-115, -20{]}$          & $-3.1 \pm 0.8$               \\
APOGEE DR17         & ${[}-115, -20{]}$          & $-3.3 \pm 2.1$               \\
APOGEE DR17         & ${[}-180, -20{]}$          & $-1.5 \pm 1.0$               \\
Hayes+20 (anchored) &         & $-4.0 \pm 0.3$               \\
Hayes+20 (internal) &          & $-1.4 \pm 1.4$               \\ \hline \hline
\end{tabular}
\end{table}

The large area coverage and number of Sgr members in the photometric sample allow us to measure the metallicity gradient along a large extension of the stream. In Figure~\ref{fig:gradient_splus}, we present the metallicity as a function of the Sgr longitude coordinate ($\Lambda_{\mathrm{SGR}}$) for the ANN (top), RF (middle), and APOGEE DR16 (bottom) samples. The APOGEE sample, previously analyzed by \citet{Hayes+2020}, is included for direct comparison. We define the trailing arm as $\Lambda_{\mathrm{SGR}} > 20^\circ$ (red) and the leading arm as $\Lambda_{\mathrm{SGR}} < -20^\circ$ (blue). The core, which is only covered by APOGEE, spans $-20^\circ \leq \Lambda_{\mathrm{SGR}} \leq 20^\circ$ (yellow). For the top and middle panels, the red and blue circles correspond to the average metallicities computed in 10$^\circ$ bins, with the error bars representing the metallicity dispersion in each bin.

We performed linear-model fits to quantify metallicity gradients along each arm. In Figure~\ref{fig:gradient_splus}, lines labeled L$1$ (trailing arm) and L$2$ (leading arm) represent linear fits using the full data range. However, the data for the trailing arm suggests a change in gradient regime near $\Lambda_{\mathrm{SGR}} = 80^\circ$. Thus, we also performed separate fits for the inner ($20^\circ \leq \Lambda_{\mathrm{SGR}} \leq 80^\circ$, line L3) and outer ($80^\circ \leq \Lambda_{\mathrm{SGR}} \leq 130^\circ$, line L4) regions. The uncertainties in the gradient measurements were estimated using a bootstrap technique; these results are summarized in Table~\ref{tab:gradient_summary}. We adopt the convention that a gradient is positive if metallicity increases outward from the core, and negative if metallicity decreases, regardless of the angular coordinate sign used in the linear fit.

We find a clear negative gradient for the leading arm: $(-2.8 \pm 1.6)$ $10^{-3}$ dex deg$^{-1}$ (ANN) and $(-3.1 \pm 0.8)$ $10^{-3}$ dex deg$^{-1}$ (RF). Conversely, the trailing arm shows negligible overall gradient when considering the full extent of our data: $(-0.4 \pm 0.2)$ $10^{-3}$ dex deg$^{-1}$ (ANN) and $(0.4 \pm 0.1)$ $10^{-3}$ dex deg$^{-1}$ (RF). However, dividing the trailing arm into inner and outer sections reveals distinct regimes. Nearer to the core ($20^\circ \leq \Lambda_{\mathrm{SGR}} \leq 80^\circ$), gradients are clearly negative: $(-3.6 \pm 0.8)$ $10^{-3}$ dex deg$^{-1}$ (ANN) and ($-3.5 \pm 0.5)$ $10^{-3}$ dex deg$^{-1}$ (RF). Beyond this region ($80^\circ \leq \Lambda_{\mathrm{SGR}} \leq 130^\circ$), the gradient reverses sign, becoming positive: $(2.2 \pm 1.8)$ $10^{-3}$ dex deg$^{-1}$ (ANN) and $(2.1 \pm 0.8)$ $10^{-3}$ dex deg$^{-1}$ (RF). Thus, the lack of an overall gradient in the trailing arm results from averaging these distinct inner and outer behaviors. Additionally, within uncertainties, gradient estimates are robust to the photometric metallicity estimation method (ANN vs. RF).

For comparison, \citet{Hayes+2020} measured internal metallicity gradients of $(-1.2 \pm 0.9)$ and $(-1.4 \pm 1.4)$ $10^{-3}$ dex deg$^{-1}$ for the trailing and leading arms, respectively (which are represented by gray dashed lines in the bottom panel of Figure~\ref{fig:gradient_splus}). The authors also present anchored gradients, which incorporate the core's metal-rich stars. In these cases, the gradients are significantly steeper: $(-2.6 \pm 0.4)$ $10^{-3}$ dex deg$^{-1}$ for the trailing arm, and $(-4.0 \pm 0.3)$ $10^{-3}$ dex deg$^{-1}$ for the leading arm, both represented by solid black lines. The internal gradients from \citet{Hayes+2020} agree marginally with our results. A more direct comparison is obtained by restricting the APOGEE DR17 data to the same area covered by S-PLUS ($20^\circ \leq \Lambda_{\mathrm{SGR}} \leq 130^\circ$ trailing; $-130^\circ \leq \Lambda_{\mathrm{SGR}} \leq -20^\circ$ leading). When recomputing the gradients considering only these regions we obtain $(1.4 \pm 2.5)$ $10^{-3}$ dex deg$^{-1}$ for the trailing, and $(-3.3 \pm 2.1)$ $10^{-3}$ dex deg$^{-1}$ for the leading arm, which are in agreement with our photometric measurements. We also computed APOGEE DR17 gradients for the full extent of the arms ($20 \leq |\Lambda_\mathrm{Sgr}| \leq 180$), which are consistent with the internal gradients measured by \citet{Hayes+2020}. Except for localized discrepancies (i.e., smaller subregions covering 40--50 degrees along the trailing arm that display more pronounced positive or negative gradients), the gradient measured for the trailing arm is negligible. This result is consistent with the reports of \citet{Muraveva+2025}: using RR Lyrae stars, these authors also observe a negligible gradient for the trailing and a negative gradient for the leading arm.

\section{Metallicity in N-body simulation}

\label{sec:nbody}

One of the main goals in studying the present-day structure of the Sgr stream is to infer the original properties of its progenitor galaxy. In particular, the observed metallicity gradient along the stream is thought to reflect the tidal stripping of a progenitor that already possessed a radial metallicity gradient. In such a scenario, the outer, more metal-poor stars, which are less tightly bound, were stripped earlier, while the more metal-rich and centrally concentrated stars remained bound longer. As a result, the present-day metallicity distribution retains a fossil imprint of the progenitor’s original gradient.

In their study, \citet{Cunningham+2024} have demonstrated that the present-day metallicity distribution along the stream is consistent with an initial radial metallicity gradient of $\sim{-0.1}$ to ${-0.2}$ dex kpc$^{-1}$. Their approach involved painting metallicities onto an N-body simulation of the stream and comparing the predicted trends with Gaia XP spectroscopic data. Here, we adopt a similar methodology, developed independently, to further explore the imprint of such gradients on present-day stream structure using a different simulation setup.

We use the N-body model of \citet{Vasiliev+2021a}, which includes the gravitational influence of the Large Magellanic Cloud and accurately reproduces the present-day morphology and kinematics of the Sgr stream. To simulate metallicity gradients, we assign metallicities to particles in the progenitor galaxy based on their initial galactocentric radius, effectively imprinting a radial [Fe/H] gradient prior to disruption. We explore a grid of 60 different gradient inputs, each uniformly sampled between $-1.0$ and $+0.2$ dex kpc$^{-1}$. In all cases, the central metallicity is fixed at [Fe/H] = $-0.5$, matching the peak of the observed metallicity distribution in the Sgr core \citep{Hayes+2020}. This approach allows us to investigate how a wide range of initial gradients would manifest in the present-day structure of the stream. We note that our model implicitly assumes that no significant star formation burst occurred in the Sgr core during its first pericentric passages, or that any such burst was at least insufficient to substantially enrich the nucleus. Starbursts associated with pericentric passages have been previously predicted in hydrodynamical simulations \citep[e.g.,][]{Amarante+2022}, but incorporating this effect would require a more sophisticated chemodynamical simulation, which lies beyond the scope of this study.

\begin{figure*}
 \includegraphics[width=\linewidth]{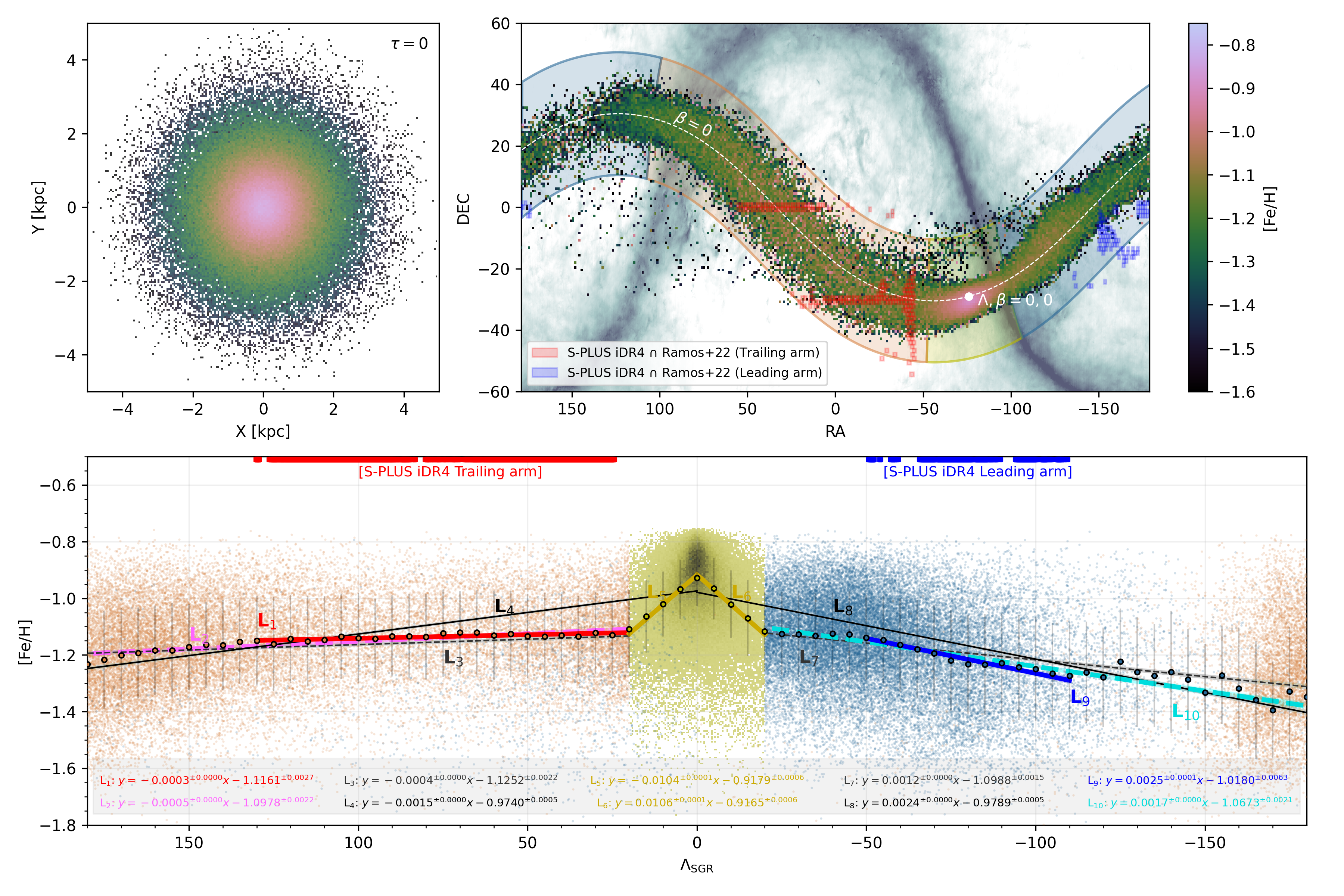}
 \caption{Illustration of the simulated metallicity gradient along the Sgr stream. Top-left: Initial spatial distribution of particles in the progenitor, color-coded by imprinted metallicity. Top-right: Present-day sky-projected positions of the same particles after tidal disruption, showing the resulting metallicity distribution. Shaded regions mark the core (yellow), trailing arm (red), and leading arm (blue). Bottom: Simulated metallicity as a function of stream longitude, $\Lambda_{\mathrm{Sgr}}$. Particles are colored by stream region, and linear fits (L1–L10) illustrate gradient estimates for different spatial selections. Red and blue horizontal top segments indicate the angular extent of the S-PLUS DR4 coverage for comparison with observations.}
 \label{fig:simulated_gradient}
\end{figure*}

Figure~\ref{fig:simulated_gradient} illustrates the methodology for a representative example from our simulation grid. In this example, metallicities were assigned to particles in the progenitor according to a radial gradient of [Fe/H] $= -0.2 \cdot R - 0.5$, where $R$ is the initial galactocentric distance in kpc. The top-left panel shows the initial configuration of the satellite, where the particles are color-coded according to their imprinted metallicity. The top-right panel presents the present-day sky-projected positions of these same particles, after tidal evolution. This panel highlights how the metallicity field is redistributed along the stream. The core region (yellow-shaded) retains the originally metal-rich stars from the center of the progenitor, while the outer, more metal-poor stars populate the leading (blue-shaded) and trailing (red-shaded) arms. This example demonstrates how an initial radial gradient can produce a measurable metallicity structure along the stream, enabling constraints on the progenitor’s internal composition. For reference, the Schlegel extinction map \citep{Schlegel+1998} is plotted in the background, and the S-PLUS coverage is plotted on top of the simulated particles as red squares for the trailing arm and blue squares for the leading arm.

The lower panel of Figure~\ref{fig:simulated_gradient} is the simulated counterpart to Figure~\ref{fig:gradient_splus}, which presents the metallicity of simulated particles as a function of their present-day stream longitude, $\Lambda_{\mathrm{Sgr}}$. These metallicities reflect the values imprinted at the progenitor’s initial configuration, mapped forward using the N-body simulation. Particles are color-coded by stream region: trailing arm ($\Lambda_{\mathrm{Sgr}} > 20^\circ$, red), leading arm ($\Lambda_{\mathrm{Sgr}} < -20^\circ$, blue), and core ($-20^\circ \leq \Lambda_{\mathrm{Sgr}} \leq 20^\circ$, yellow). While throughout the paper we define the longitude range as being wrapped within $[-180^\circ, +180^\circ]$, this panel includes stars beyond that range, as the simulation tracks particles that have completed additional orbital loops. To guide visual comparison with the observational data, we overplot red and blue horizontal segments marking the S-PLUS DR4 spatial coverage along the trailing and leading arms, respectively. Finally, to account for the variation in how metallicity gradients are reported in the literature, we include multiple linear fits across different segments of the stream, enabling direct comparison with observational analyses.

The various linear fits shown in the lower panel of Figure~\ref{fig:simulated_gradient} illustrate how the recovered metallicity gradients depend strongly on the region of the stream considered. The steepest gradients are observed near the core, as indicated by the yellow solid lines (L5 and L6), reflecting the sharp transition in metallicity between the inner and outer parts of the stream. The black solid lines (L4 for $\Lambda_{\mathrm{Sgr}} > 0$ and L8 for $\Lambda_{\mathrm{Sgr}} < 0$) show fits across the full trailing and leading arms, including the core. These also exhibit pronounced gradients, but this is primarily driven by the concentration of metal-rich stars near the remnant, underscoring the importance of excluding the core when measuring gradients along the stream. 

When we exclude the central region ($-20^\circ < \Lambda_{\mathrm{Sgr}} < 20^\circ$), the dashed black lines (L3 and L7) show significantly shallower gradients, with the trailing arm gradient becoming essentially negligible. We refine this analysis further by removing stars that have completed additional orbital wraps: the dashed pink (L2) and dashed cyan (L10) lines correspond to fits restricted to unwrapped particles in the trailing and leading arms, respectively. While the trailing arm still shows no significant gradient, the leading arm gradient becomes more pronounced once contamination is removed. Lastly, to enable direct comparison with the S-PLUS DR4 observations, we fit the metallicity gradient over the same angular extent sampled by the data. These are shown as solid red (L1, trailing arm) and solid blue (L9, leading arm) lines. The simulated stream reproduces the key observational result: a negligible metallicity gradient in the trailing arm and a significantly steeper negative gradient in the leading arm.

Overall, the simulation successfully reproduces two key observational features of the Sgr stream. First, it naturally yields a metallicity offset between the arms, with the leading arm being systematically more metal-poor than the trailing arm—consistent with numerous spectroscopic studies \citep[e.g.,][]{Carlin2018sgr, Hayes+2020, Limberg+2023}. This reflects the earlier stripping of the outer, metal-poor stellar populations into the leading arm, while more metal-rich stars, initially concentrated near the core, dominate the trailing debris. Second, the simulation recovers the distinct behavior of the metallicity gradients in each arm: negligible in the trailing arm and significant in the leading arm, in agreement with recent results \citep[e.g.,][]{Cunningham+2024, Muraveva+2025}. These outcomes reinforce the interpretation that the present-day chemical structure of the stream retains measurable signatures of the progenitor’s internal metallicity profile and stripping history.

\begin{figure*}
 \includegraphics[width=\linewidth]{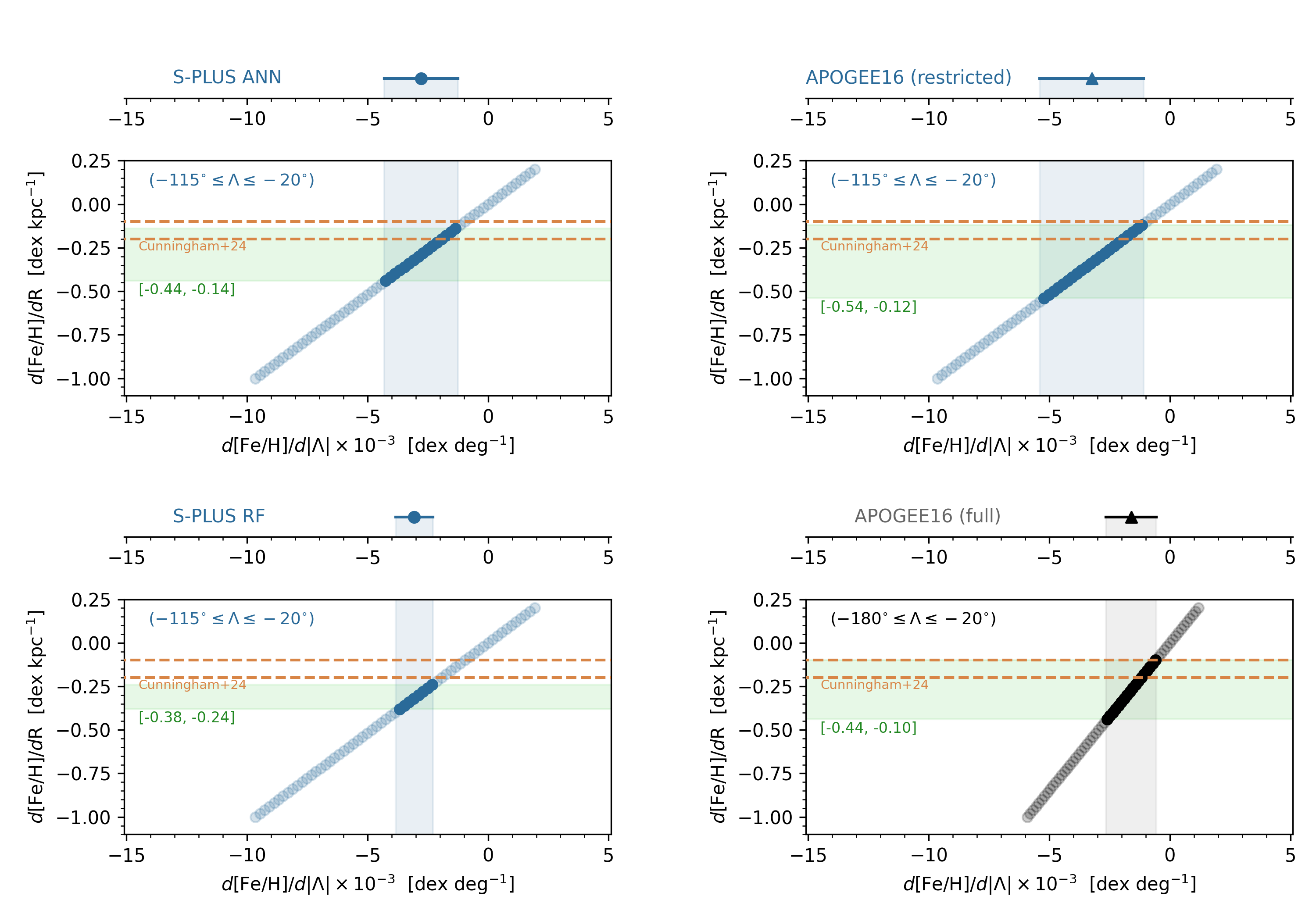}
 \caption{Simulated present-day metallicity gradients along the leading arm of the Sagittarius stream (x-axis) as a function of the simulated progenitor radial gradient (y-axis), for the estimations from S-PLUS ANN (top-left), S-PLUS RF (bottom-left), restricted Hayes+20 sample (top-right), and full Hayes+20 sample (bottom-right). The restricted Hayes sample covers the same region as S-PLUS, while the full sample spans the whole leading arm. In each panel, the top axis shows the observed gradient, with errors represented by the standard deviation of the bootstrap linear fitting. The estimated observed gradient range is used to constrain the models that result in such gradients. The constrained models are highlighted and represented by the shaded green region. The interval estimated by \citet{Cunningham+2024} is shown as a red dashed line for comparison.}
 \label{fig:predicted_gradient}
\end{figure*}

\subsection{Constraining the Sgr progenitor radial metallicity gradient}

In order to quantitatively compare our photometric measurements with model predictions, we focus on the slope of the present-day metallicity gradient along the stream in two specific regions: (i) the fraction of the leading arm covered by S-PLUS ($-20^{\circ} > \Lambda_{\mathrm{Sgr}} \geq -115^{\circ}$), and (ii) the full leading arm ($-180^{\circ} < \Lambda_{\mathrm{Sgr}} < -20^{\circ}$). These choices allow for a direct comparison between observations and simulations in both the S-PLUS (case i) and spectroscopic samples (both cases). For this analysis, the trailing arm was not considered, as both the simulations and observations indicate a negligible present-day gradient, which offers no possibility of constraining the original properties of the progenitor.

This comparison is shown in Figure~\ref{fig:predicted_gradient}. The upper sub-axis in each panel presents the measured value for each case: S-PLUS ANN (top-left), S-PLUS RF (bottom-left), APOGEE DR17, restricted to the S-PLUS DR4 footprint (top-right), and APOGEE DR17 covering the full extent of the leading arm (bottom-right). Within each panel, the vertical shaded region indicates the observationally constrained present-day metallicity gradient. The circles represent the predicted present-day gradients (x-axis) obtained from simulations with different assumed initial radial metallicity gradients (y-axis). The green horizontal shaded region highlights the range of initial conditions that reproduces the observed present-day values (considering the uncertainties). For comparison, the dashed red lines correspond to the values constrained by \citet{Cunningham+2024}. Notably, we observe a linear relationship between the slope of the progenitor's radial gradient, and the slope of the present-day gradient along the leading arm.

Considering all four cases, the RF-based photometric metallicities for the S-PLUS DR4 are the ones that provide the tightest constraint on the progenitor's properties, resulting in an initial radial metallicity gradient in the range $[-0.38, -0.24]$ dex kpc$^{-1}$. In contrast, the spectroscopic sample, even when considering the full extent of the leading arm, restricts the initial $[-0.42, -0.10]$ dex kpc$^{-1}$. Nevertheless, all four determinations are consistent within their respective uncertainties, which reinforces the robustness of the overall picture. This result highlights the strength of photometric metallicities for this type of study, despite being less precise for individual stars. Overall, the much larger S-PLUS sample size allows for a tighter constraint on the progenitor's original gradient than what is achieved with spectroscopy alone.

\begin{figure*}
 \includegraphics[width=\linewidth]{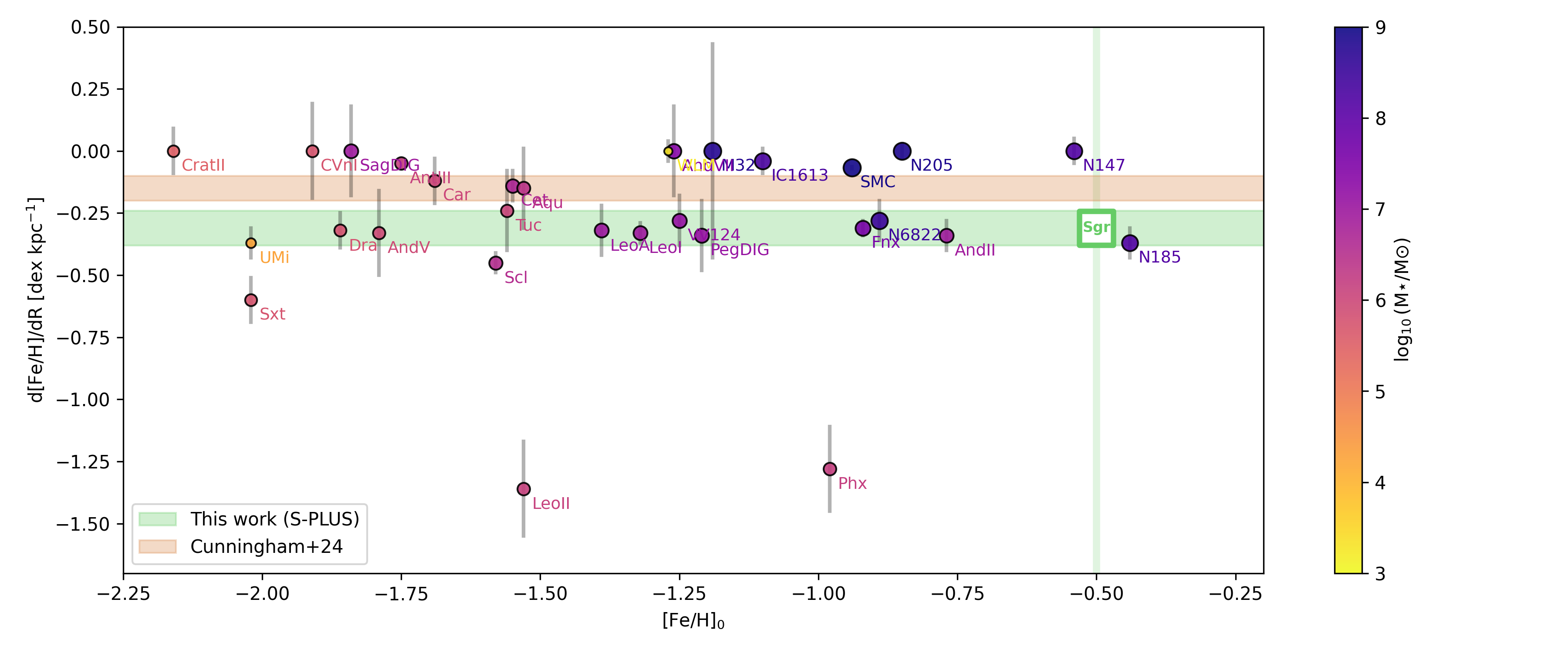}
 \caption{Radial metallicity gradients of Local Group dwarf galaxies as a function of their central metallicity. Symbol size and color reflect the total stellar mass from \citet{McConnachie2012}. The green shaded region indicates the gradient interval inferred for the Sgr progenitor based on photometric data. The vertical green line marks its central metallicity from \citet{Hayes+2020}. The red shaded area corresponds to the gradient range reported by \citet{Cunningham+2024}.}
 \label{fig:comparison_gradient}
\end{figure*}

Notably, these constraints diverge from those reported in prior work, particularly by \citet{Cunningham+2024}. While our analysis constrains the original radial metallicity gradient of the Sagittarius progenitor to lie within the range $-0.38$ to $-0.24$ dex kpc$^{-1}$, these authors estimate a significantly shallower gradient of $\sim-$0.1 to $-0.2$ dex kpc$^{-1}$. 

This discrepancy could be explained by the differences in the N-body simulations that produce the reference model. While our analysis is based on the N-body simulation of \citet{Vasiliev+2021a}, \citet{Cunningham+2024} developed their own simulation (based on the same galactic dynamics library, \texttt{Agama}, \citealp{Vasiliev2019}). As noted by \citet{Cunningham+2024}, the morphology of the present-day state of the simulation is dependent on the orbital phase, which may not match the current real orbital phase of the stream. Nevertheless,  we employ a robust model-constraining approach to link present-day gradients to the progenitor's initial conditions, which can be reproduced in future studies, with better-constrained modeling of the stream, to either support or refute these results. 

We also note that we assumed a linear gradient, with a central metallicity of $-0.5$ dex for all the models considered in our grid, which could also be non-representative of the true original radial gradient of the Sgr progenitor. We encourage future work to explore a comparative analysis of the predicted progenitor gradient using multiple sets of N-body simulations and considering a more extended set of profiles for the initial metallicity gradient.

\subsection{Comparison with Local Group Dwarf Galaxies}

We compare our inferred metallicity gradient with radial metallicity gradients measured in dwarf galaxies across the Local Group. Figure~\ref{fig:comparison_gradient} shows the radial gradient versus the central metallicity for a sample of dwarf galaxies from the compilation of \citet{Taibi+2022}. The size and color of each point encode the total stellar mass of the system \citep[from][]{McConnachie2012}.

Our results for the Sagittarius progenitor are shown as horizontal green bands, corresponding to the constraints obtained the RF photometric metallicities. For reference, the central metallicity of Sagittarius ([Fe/H]$_0 \approx -0.5$ dex) is marked as a vertical green line, based on the APOGEE core measurements from \citet{Hayes+2020}. The red shaded region represents the interval inferred by \citet{Cunningham+2024}.

As shown in Figure~\ref{fig:comparison_gradient}, the radial metallicity gradient we infer for the Sagittarius progenitor is consistent with those observed in other dwarf galaxies of comparable mass. Although the estimate by \citet{Cunningham+2024} is shallower than ours, their range also overlaps with several known systems, including some relatively low-mass dwarfs. Consequently, neither result appears anomalous in the broader context of Local Group galaxy properties.

The strongest correlation in this sample appears between stellar mass and central metallicity: more massive dwarf galaxies tend to have more metal-rich cores. Sagittarius, with its relatively high central [Fe/H], aligns more closely with systems such as NGC 185 and NGC 147, suggesting its progenitor likely had a stellar mass given by $\log_{10} (M_\star / M_\odot) \sim 8.2$. For instance, \citet{Vasiliev+Belokurov2020} estimated a stellar mass of $\sim 1 \cdot 10^8$ M$_\odot$ (out of a total mass of $\sim 4 \cdot 10^8$ M$_\odot$) for the present-day bound Sgr remnant, slightly lower than the masses of NGC 147 and NGC 185, as expected given the extensive tidal stripping experienced by Sagittarius.

\section{Conclusions} \label{sec:conclusions}

In this work, we investigated whether the present-day metallicity distribution along the Sagittarius stream retains imprints of an intrinsic radial metallicity gradient in its progenitor galaxy. Our goals were to characterize the metallicity distribution and gradient along the stream using photometric metallicity estimates, and to constrain the original gradient in the Sgr dSph by comparison with predictions from $N$-body simulations for a set of initial metallicity profiles.

Our analysis is based on a crossmatch between S-PLUS DR4 photometry and the Gaia DR3 Sagittarius stream catalog from \citet{Ramos+2022}, resulting in over 17,500 candidate stream members with estimated photometric metallicities. We considered metallicities estimated from two machine learning models: Random Forest and Artificial Neural Networks. Additionally, we also considered APOGEE DR17 data to enable direct comparison with previous spectroscopic results.

We find a clear metallicity offset between the two arms of the Sagittarius stream, with the leading arm being 0.10–0.20 dex more metal-poor than the trailing arm. The result is consistent across both RF and ANN photometric estimates and aligns with previous spectroscopic studies \citep[e.g.][]{Carlin2018sgr, Hayes+2020, Limberg+2023}. This supports the scenario in which the leading arm originates from earlier stripping of the outer, metal-poorer regions of the progenitor.

We detected a clear negative metallicity gradient along the leading arm. In contrast, the trailing arm shows no significant gradient when considered as a whole. However, a more detailed analysis reveals the possible existence of multiple regimes within the trailing arm. For instance, we observe a negative gradient in the inner trailing region ($20^\circ \leq \Lambda_{\mathrm{SGR}} \leq 80^\circ$) and a positive gradient in the outer region, indicating a complex structure which could be related to multiple stripping episodes. These trends are consistent across both ANN and RF photometric estimates, and also to spectroscopic results from APOGEE, when restricted to the same spatial coverage. Nevertheless, our findings confirm that the metallicity varies systematically along the stream. When considering the same sampling restrictions, our results are consistent with other studies \citep[e.g.][]{Chou+2007, Monaco+2007, Hayes+2020, Muraveva+2025}).

In order to interpret the observed metallicity trends, we employed $N$-body simulations of the Sagittarius stream based on the model from \citet{Vasiliev+2021a}. First, we assigned metallicities to the simulated particles as a function of their initial galactocentric radius. This process was repeated for a set of 60 different gradient slopes. Then, we used the present-day predictions of these models to compare the resulting metallicity distributions to our observational results. We find that the simulated stream reproduces both the metallicity offset between the arms and the contrasting gradient behavior: steep in the leading arm, and negligible in the trailing, when restricted to the same spatial coverage as S-PLUS. These results confirm that metallicity gradients observed along the stream can be used to constrain the internal structure of the progenitor galaxy.

We compared the observed metallicity gradient to the predictions for the N-body simulations for different initial slopes. From this data, we were able to constrain the original radial metallicity gradient of the Sgr progenitor to lie between $-0.38$ and $-0.24$ dex kpc$^{-1}$ based on the S-PLUS data, and between $-0.42$ and $-0.10$ dex kpc$^{-1}$ from the APOGEE sample. The models are constrained mostly by the present-day slope observed in the leading arm, while the trailing arm gradient is largely insensitive to the progenitor’s initial profile. Notably, \citet{Cunningham+2024} report a much shallower metallicity gradient. The reason for this could be the differences in the employed N-body simulation between the studies. We emphasize that more work needs to be done in this field to better understand the discrepancy between our results.

Finally, we compared the inferred metallicity gradient of the Sagittarius progenitor with those measured in other dwarf galaxies across the Local Group. Our results are consistent with the range of gradients observed in systems of similar mass and central metallicity. While the gradient estimated by \citet{Cunningham+2024} is shallower, their values are still consistent with the diversity seen among Local Group dwarfs. The central metallicity of Sagittarius ([Fe/H]$_0 \approx -0.5$ dex) and the inferred gradient are similar to those observed in more massive dwarf galaxies such as NGC 147 and NGC 185, suggesting a stellar mass for the progenitor near $\sim 1.5 \cdot 10^8$ M$_\odot$, which is in agreement with previous mass estimates for Sgr \citep{Vasiliev+Belokurov2020}.

In summary, our results show that photometric surveys with metallicity-sensitive narrow-band filters, such as S-PLUS, are powerful tools in reconstructing the chemical and dynamical histories of Galactic structures like the Sagittarius stream. Even with the limited coverage of S-PLUS DR4, which covers just a third of the area of its planned footprint, we were able to recover robust metallicity gradients that allowed us to constrain key properties of the progenitor galaxy. Future S-PLUS data releases will cover an extended area, providing better coverage along and across the stream, which will allow the subsequent studies to also explore additional signatures such as vertical gradients perpendicular to the stream. Finally, while our conclusions are supported by independent spectroscopic data and realistic $N$-body modeling, future work should also test alternative dynamical histories and explore different assumptions about the gradient shape and central metallicity of the Sgr progenitor galaxy.

\begin{acknowledgments}

The S-PLUS project, including the T80-South robotic telescope and the S-PLUS scientific survey, was founded as a partnership between the Fundação de Amparo à Pesquisa do Estado de São Paulo (FAPESP), the Observatório Nacional (ON), the Federal University of Sergipe (UFS), and the Federal University of Santa Catarina (UFSC), with important financial and practical contributions from other collaborating institutes in Brazil, Chile (Universidad de La Serena), and Spain (Centro de Estudios de Física del Cosmos de Aragón, CEFCA). We further acknowledge financial support from the São Paulo Research Foundation (FAPESP) grant 2019/263492-3, the Brazilian National Research Council (CNPq), the Coordination for the Improvement of Higher Education Personnel (CAPES), the Carlos Chagas Filho Rio de Janeiro State Research Foundation (FAPERJ), and the Brazilian Innovation Agency (FINEP). F.A.-F. acknowledges support from FAPESP grants 2024/00822-5 and 2024/22842-8. G.L. acknowledges support from KICP/UChicago through a KICP Postdoctoral Fellowship. J.A. is supported by the National Natural Science Foundation of China under grant Nos. 12233001, 12533004, by  the National Key R\&D Program of China under grant No. 2024YFA1611602, by a Shanghai Natural Science Research Grant (24ZR1491200), by the ``111'' project of the Ministry of Education under grant No. B20019, by the China Manned Space Program with grant Nos. CMS-CSST-2025-A08, CMS-CSST-2025-A09 and CMS-CSST-2025-A11, and in part by Office of Science and Technology, Shanghai Municipal Government (grant Nos. 24DX1400100, ZJ2023-ZD-001). G.B. acknowledges support from FAPESP grant 2024/02676-6. M.B.F. acknowledges financial support from the National Council for Scientific and Technological Development (CNPq) Brazil (grant number: 307711/2022-6). The work of V.M.P. is supported by NOIRLab, which is managed by the Association of Universities for Research in Astronomy (AURA) under a cooperative agreement with the U.S. National Science Foundation. ARS acknowledge support by FAPESP through proc. 2025/19654-8 and the Brazilian National Council for Scientific and Technological Development (CNPq) grant 192390/2025-2. K.M.D. thanks the support of the Serrapilheira Institute (grant Serra-1709-17357) as well as that of the Brazilian National Research Council (CNPq grant 308584/2022-8) and of the Rio de Janeiro Research Foundation (FAPERJ grant E-32/200.952/2022), Brasil. S.R. also thanks partial financial support from FAPESP (Proc. 2020/15245-2), CNPq (Proc. 303816/2022-8), and CAPES.
 
\end{acknowledgments}




%
\facilities{S-PLUS Survey (CTIO - T80-South telescope)}

\software{Astropy \citep{astropy:2022},
          NumPy \citep{harris2020array},
          SciPy \citep{virtanen2020scipy},
          Matplotlib \citep{hunter2007matplotlib},
          pandas \citep{mckinney2010data},
          scikit-learn \citep{scikit-learn},
          Cartopy \citep{cartopy}}


\appendix
\section{Validation of Photometric Metallicities}\label{apend:photmet}

\begin{figure*}
 \includegraphics[width=\linewidth]{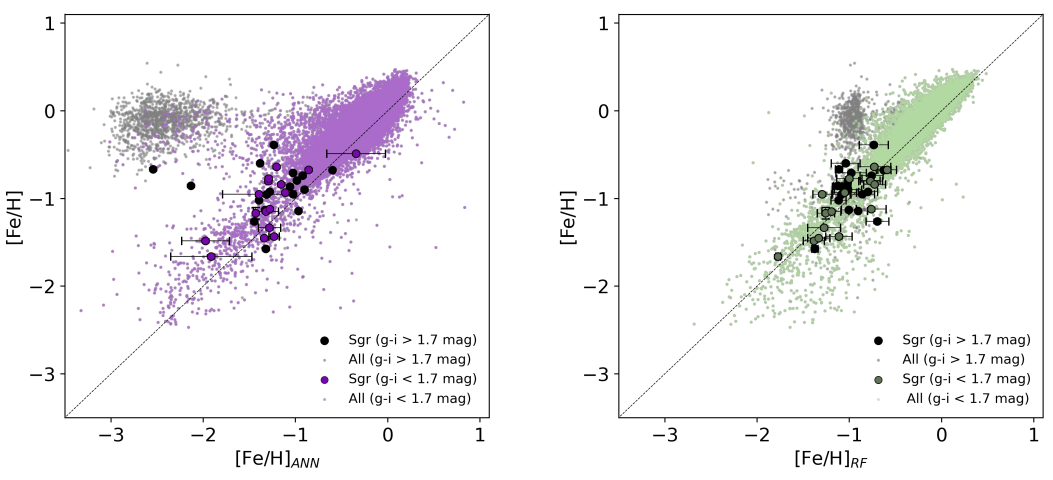}
 \caption{Comparison between S-PLUS photometric metallicities and APOGEE DR17 spectroscopic metallicities for the ANN (left) and RF (right) samples. The full sample is shown as dots, while the stars belonging to our Sgr sample are shown as circles. The colored markers correspond to the stars with reliable parameters $(g-i) < 1.7$, while colder stars are shown as grey/black markers.}
 \label{fig:comp_feh_apogee}
\end{figure*}

In this Appendix we compare the S-PLUS photometric metallicities with those from APOGEE DR17, for a total of 12{,}532 stars in the crossmatch with the Random Forest (RF) sample, and 14{,}180 stars in the Artificial Neural Network (ANN) sample. This comparison is shown in Figure \ref{fig:comp_feh_apogee} for the full crossmatch (colored dots), and for the stars both in the APOGEE and Sagittarius sample (46 stars, for both methods) described in Section\ref{sec:data} (colored circles).

Overall, we observe a good agreement between our photometric metallicities and the APOGEE DR17 spectroscopic metallicities. Nevertheless, we can see one 'blob' in each plot, centered around [Fe/H]$_\mathrm{ANN}$ $\sim -2.5$ and [Fe/H] $\sim 0.0$, for the ANN method, and around [Fe/H]$_\mathrm{RF}$ $\sim -1.0$ and [Fe/H] $\sim 0.0$ for the RF sample. We identify these as cold stars ($(g-i)$ > 1.7), which are not covered in the RF and ANN training samples.

For the ANN method, the average difference between photometric metallicities is 0.34~dex, with a standard deviation of 0.65~dex. When removing the stars with $(g-i) > 1.7$, the average difference improves to 0.16~dex, and standard deviation to 0.28~dex.

The situation is similar in the RF method: when removing stars with $(g-i) > 1.7$, the average difference improves from 0.12~dex to 0.08~dex, while the standard deviation improves from 0.25~dex to 0.17~dex.

\bibliography{sample701}{}
\bibliographystyle{aasjournalv7}



\end{document}